\documentclass[aps,preprint,superscriptaddress,showpacs]{revtex4}
\usepackage{graphicx}

\usepackage{subfigure}

\begin{document}

\author{Fabrizio Capuani}

\affiliation{FOM Institute for Atomic and Molecular Physics (AMOLF), Kruislaan 407, 1098 SJ Amsterdam, The Netherlands}

\email{capuani@amolf.nl, frenkel@amolf.nl}

\author{Ignacio Pagonabarraga}

\affiliation{Departament de F\'{\i}sica Fonamental, C. Mart\'{\i} i Franqu\'es 1, 08028 Barcelona, Spain}

\email{ipagonabrraga@ub.edu}

\author{Daan Frenkel}

\affiliation{FOM Institute for Atomic and Molecular Physics (AMOLF), Kruislaan 407, 1098 SJ Amsterdam, The Netherlands}

\email{frenkel@amolf.nl}

\title{Discrete solution of the electrokinetic equations}


\begin{abstract}
We present a robust scheme for solving the electrokinetic
equations. This goal is achieved by combining the
lattice-Boltzmann method (LB) with a discrete solution of the
convection-diffusion equation for the different charged and
neutral species that compose the fluid. The method is based on
identifying the elementary fluxes between nodes, which ensures the
absence of spurious fluxes in equilibrium. We show how the model
is suitable to study electro-osmotic flows. As an illustration, we
show that, by introducing appropriate dynamic rules in the
presence of solid interfaces, we can compute the sedimentation
velocity (and hence the sedimentation potential) of a charged
sphere. Our approach does not assume linearization of the
Poisson-Boltzmann equation and allows us for a wide variation
of the Peclet number.
\end{abstract}

\maketitle

\section{Introduction}

The study of the dynamics of suspensions of charged particles is
interesting both because of the subtle physics underlying many
electrokinetic phenomena and because of the practical relevance of
such phenomena for the behavior of many synthetic and biological
complex fluids~\cite{holm,pincus}. In particular, electrokinetic
effects can be used to control the transport of charged and
uncharged molecules and colloids, using electrophoresis,
electro-osmosis, and related phenomena~\cite{Probstein}. As
micro-fluidic devices become ever more prevalent, there are an
increasing number of applications of electro-viscous phenomena
that can be exploited to selectively transport material in devices
with mesoscopic dimensions~\cite{stone}.

In virtually all cases of practical interest, electroviscous
phenomena occur in confined systems of a rather complex geometry.
This makes it virtually hopeless to apply purely analytical
modeling techniques. But also from a molecular-simulation point of view
electroviscous effects present a formidable challenge. First of
all, the systems under consideration always contain at least three
components; namely a solvent plus two (oppositely charged) species. Then,
there is the problem that the physical properties of the systems
of interest are determined by a number of potentially different
length scales (the ionic radius, the Bjerrum length, the
Debye-H\"{u}ckel screening length and the characteristic size of
the channels in which transport takes place). As a result, fully
atomistic modeling techniques become prohibitively expensive for
all but the simplest problems. Conversely, standard discretization
of the macroscopic transport equations is ill suited to deal with
the statistical mechanics of charge distributions in ionic
liquids, even apart from the fact that such techniques are often
ill-equipped to deal with complex boundary conditions.

In this context, application of mesoscopic ("coarse-grained")
models to the study of electrokinetic phenomena in complex fluids
seem to offer a powerful alternative approach. Such models can be
formulated either by introducing effective forces with dissipative
and random components, as in the case of dissipative particle
dynamics (DPD)~\cite{groot}, or by starting from simplified
kinetic equations, as is the case with the lattice-Boltzmann
method (LB).

The problem with the DPD approach is that it necessarily
introduces an additional length scale (the effective size of the
charged particles). This size should be much smaller than the
Debye screening length, because otherwise real charge-ordering
effects are obscured by spurious structural correlations; hence, a
proper separation of length scales may be difficult to
achieve.
A Lattice-Boltzmann model for
electroviscous effect was proposed by Warren~\cite{PBWarren}. In
this model, the densities of the (charged)  solutes are treated as
passive scalar fields. Forces on fluid element are mediated by
these scalar fields. A different approach was followed in
Ref.~\cite{Juergen}, where solvent and solutes are treated on the
same footing (namely as separate species). This method was then
extended to couple the dynamics of charged colloids to that of the
electrolyte solution. As we shall discuss below, both approaches
have practical drawbacks that relate to the mixing of discrete and
continuum descriptions.

The LB model that we introduce below appears at first sight rather
similar to the model proposed by Warren. However, the underlying
philosophy is rather different. We propose to consider the {\em
fluxes}\ between connected nodes as the basic physical quantities
that determine the evolution of local densities. Such a
formulation ensures local mass conservation, does not rely on
fluxes or gradients computed at the lattice nodes (which
constitutes a source of error in other models due to the need to
approximate them on a lattice), and by choosing a symmetric
formulation for the link fluxes in terms of the  nodes that are
affected, we can recover the proper equilibrium without spurious
fluxes. Our model relies on an LB formulation for mixtures. Hence,
the improvements of the formulation based on link fluxes will
overcome some of the limitation of previous LB models for mixtures
based on gradient expansions of a free energy~\cite{swift}.

The method described is very flexible, and in particular general
boundary conditions are easily implemented. This feature makes
also the proposed formulation attractive, since it avoids problems
related to mass and charge conservation at fluid-solid interfaces,
an artifact that has plagued previous LB implementations. It is
then possible to model the dynamics of colloidal particles and
polyelectrolytes in solution.  The electrostatic interaction
between them is derived from the charge distribution in the fluid.
Hence, we do not need to assume any specific form for the
interaction between charged colloids, or between monomers in a
polyelectrolyte. Electro-osmosis, the sedimentation potential,
electrophoresis or other electrokinetic phenomena can be easily
treated within the model. In this paper we consider the first two
to illustrate the capabilities of the method.

The electrolyte is treated at the Poisson-Boltzmann level. We are
not restricted to the linearized Debye-H\"{u}ckel regime and can
study the electrokinetic effects at high charge densities, being
only limited by ionic  condensation (as happens for example in
cylinders).  The model we introduce will miss effects due to charge
correlations.

The remainder of this paper is organized as follows. In Section 2
we describe the hydrodynamics of fluid mixtures to set the general
background. In Section 3 we describe the proposed numerical method
and subsequently, in Section 4, we discuss how to model general
solid interfaces within this lattice model.  Section 5 focuses on
the special case of interest to treat electrokinetic phenomena. In
Section 6 we validate the method by analyzing different situations
of interest, including electro-osmosis and sedimentation.

\section{Hydrodynamic description of non-ideal fluid mixtures}
In some respects, the dynamics of electrolytes at hydrodynamic
scales is analogous to that of multicomponent mixtures. The
simplest electrolyte model consists of two ionic species and a
neutral solvent. In order to provide the general framework for the
description of electrolyte dynamics, we first briefly review the
dynamics of mixtures on hydrodynamic length and time scales. As in
all hydrodynamic descriptions, the starting point of any
discussion are the laws of conservation of mass and momentum.

\subsection{ Mass conservation}
Every species of the fluid mixture satisfies the usual mass
conservation law:
\begin{equation}
\label{Single species mass conservation}
\frac{\partial \rho _{k}}{\partial t}+\nabla \cdot \rho _{k}{\bf v}_{k}=0,
\end{equation}
where \( {\bf v}_{k} \) is the velocity and \( \rho _{k} \) the
density distribution of the species labeled by \( k \).
The total density, \(\rho = \sum_k \rho_k\), is also conserved, and satisfies an equation analogous to
Eq.~(\ref{Single species mass conservation}) with respect to the barycentric velocity
 $\rho{\bf v}=\sum_k\rho_k{\bf v}_k$, which describes the evolution of a fluid element.
If we refer the motion of all species to this common velocity,
then Eq.~(\ref{Single species mass conservation}) can be expressed as
\begin{equation}
\label{Convective Diffusion}
\frac{\partial \rho _{k}}{\partial t}+{\bf \nabla }\cdot \rho _{k}{\bf v}=-\nabla \cdot {\bf j}_{k},
\end{equation}
where we have introduced the relative current of species $k$,
${\bf j}_k=\rho_k({\bf v}_k-{\bf v})$, which accounts for all
dynamical effects arising from the mismatch in velocities between
the different species. On very short time scales, such  currents
are controlled by friction relaxation. However, for mixtures
composed of molecular constituents (as is usually the case in
electrolytes), the inertial time scale is extremely small; hence
the relative current can be assumed proportional to a
thermodynamic driving force, which is proportional to the gradient
of the chemical potential. As a result, the relative current of
species $i$ becomes diffusive and can be expressed as~\cite{dGM}
\begin{equation}
\label{Diffusive current}
{\bf j}_k=-\sum_i D_{ik}\rho _{k}{\nabla }\beta \mu _{k},
\end{equation}
where \( \beta  \) is $1/k_BT$, with $k_B$ the Boltzmann constant
and $1/T$ the inverse temperature. \( \beta \mu _{k}=\log \rho
_{k} +\beta \mu _{k}^{ex} \) is the chemical potential decomposed
in an ideal and excess part, while $D_{ik}$ corresponds to the
diffusion coefficient that determines the flux of species $i$
induced by spatial variations in the chemical potential of species
$k$. For the sake of simplicity, we concentrate on the case where
cross diffusion is neglected, and hence $D_{ik}=D_i\delta_{ik}$.
By substituting the chemical potential in Eq.~(\ref{Diffusive
current}), we can then express mass conservation in the form of a
set of convection-diffusion equations, expressing the two
mechanisms that control density evolution for each species,
\begin{equation}
\label{Smoluchowski}
\frac{\partial \rho _{k}}{\partial t}+\nabla \cdot \rho _{k}{\bf v}=\nabla \cdot D_{k}\left[ \nabla \rho _{k}+\rho _{k}\nabla \beta \mu _{k}^{ex}\right] .
\end{equation}

\subsection{Momentum conservation}
Next, we consider momentum conservation.  On the same length and
time scales, momentum conservation implies that the barycentric
velocity follows the Navier-Stokes equation,
\begin{equation}
\label{General Navier-Stokes}
\frac{\partial}{\partial t} \rho {\bf v}+\nabla\cdot\rho{\bf v}{\bf v}=
\eta \nabla ^{2} {\bf v}+\xi\nabla \left( \nabla\cdot{\bf v} \right) -\nabla p+{\bf F}^{ext},
\end{equation}
where \( \eta  \) and $\xi$ are  the shear and bulk fluid viscosities, respectively,
while \( {\bf F}^{ext} \) is the external force acting on a fluid element.
 The effect of the interactions among the different species enters as a net force expressed as the gradient
of the local pressure, $p$. In the presence of spatial gradients,
the pressure has in general a tensorial character, and can be
derived from the free energy of the system. However, for ideal
electrolytes, the local pressure can always be expressed as a
scalar. Hence, for the  sake of simplicity we
will consider this situation in what follows.
 As a result, we only need to input the free energy of the mixture to determine both the
pressure and chemical potentials. Specifically, if we know the free energy per unit volume
 $\beta f({\bf r})=\sum_k\rho_k[\log(\Lambda^3 \rho_k)-1]+\beta f^{ex}$, then
\begin{eqnarray}
\beta \mu_k&=&=\log\rho_k+\beta\mu^{ex}.\nonumber\\
\beta p&=&\sum_k\rho_k \beta \mu_k-\beta f=
                   \sum_k\left(\rho_k+\rho_k\mu_k^{ex}\right)-\beta f^{ex},
\end{eqnarray}
where the free energy, the chemical potential and the pressure are
position dependent. The first term of the
pressure corresponds to the ideal-gas contribution, \(\beta
p^{id}=\sum_k\rho_k\) while the other two contain all the
information of the interactions among the fluid species. If there
is one  majority neutral component, which only contributes to the
ideal part of the pressure, then the excess component of the
pressure can be identified as the osmotic pressure of the mixture.
In general, the pressure gradient follows from Gibbs-Duhem,
\begin{equation}
\label{eq:gibbsduhem}
\beta \nabla p=\sum_k\rho_k\beta\nabla\mu_k=\sum_k\left( \nabla\rho_k+\rho_k\beta\nabla\mu_k^{ex} \right).
\end{equation}
and  acts as a force. We will use this interpretation in the LB
implementation discussed in the next section.

Using the last expression for the pressure gradient, the Navier-Stokes equation reads
\begin{equation}
  \frac{\partial}{\partial t}\rho{\bf v}+\nabla\cdot\rho{\bf v}{\bf v}=\eta\nabla^2{\bf v}+\xi\nabla\nabla \cdot {\bf v}
-\nabla p^{id}-\sum_k\rho_k\nabla\beta\mu^{ex}+{\bf F}^{ext}.
\end{equation}

\section{Numerical lattice method}
\label{section:model} We propose a model that combines a
description of momentum dynamics based on lattice-Boltzmann, with
a numerical description of the convection-diffusion equation.
Quantities are defined on the nodes of a lattice, \({\bf r}\), and
time evolves in discrete  time steps. The lattice is prescribed by
specifying its connectivity. The connections of each node are
determined by specifying the set of allowed velocities, \({\bf
c}_i\), where the sub-index $i$ runs over all the allowed
velocities. Then, each node ${\bf r}$ is connected to the nodes
${\bf r}+{\bf c}_i$.

\subsection{Diffusion model \label{Section: smolu solution}}
For convenience, let us rewrite the convection-diffusion equation,
Eq.~(\ref{Smoluchowski}), in the form
\begin{equation}
\frac{\partial}{\partial t}\rho _{k}+\nabla\cdot \rho_{k}{\bf v} =  -\nabla \cdot {\bf j}_{k},
\label{Smolu solution}
\end{equation}
where the diffusive flux is
\begin{equation}
\label{flux}
{\bf j}_{k}=-D_{k}\left( \nabla\rho_{k}+\rho_{k}\nabla\beta \mu ^{ex}_{k}\right) .
\end{equation}
For the sake of clarity, we  discuss separately the change in
density of the species \( k \) due to diffusion and to advection.
The total change in time of the density is simply the sum of the
two contributions.
\subsubsection{Diffusion} \label{Sec: Diffusion}
Let us assume for the time being that the mixture diffuses in a  fluid at rest. Eq.~(\ref{Smolu solution}) then becomes
\[
\frac{\partial}{ \partial t}\rho _{k}=-\nabla \cdot {\bf j}_{k}.
\]
Integrating both sides of this equation over a volume \( V_{0} \)
and using the Green's formula \( \int _{V_{0}}\nabla \cdot {\bf j}dV
= \oint _{A_{0}}{\bf j}\cdot \widehat{{\bf n}}dA \),
we obtain:
\begin{equation}
\label{"green"}
\frac{\partial }{\partial t}\int_{V_{0}}\rho _{k}dV=-\oint _{A_{0}}{\bf j}_{k}\cdot \widehat{{\bf n}}dA,
\end{equation}
 where \( \widehat{{\bf n}} \) is the outward unity vector normal to the surface, $A_0$, enclosing the
volume $V_0$.

As we have pointed out previously, we will consider densities defined on nodes of a
lattice and the time evolution  evolves at constant time steps. In this case, we can
identify the volume $V_0$ with the volume associated to that node, and $A_0$
is related to the connectivity of the lattice nodes.
Then, Eq.~(\ref{"green"}) states that the change of the total number of particles
enclosed in the volume corresponding to  node ${\bf r}$ equals the sum of the outward fluxes.
Such fluxes can only  take place by mass transport to the neighboring nodes that are
connected to the central node, according to the structure of the predetermined lattice connectivity.
Hence,
\begin{equation}
\label{discrete smolu}
n_{k}\left( {\bf r}, t+1\right) -n_{k}\left( {\bf r}, t\right)=-A_0\sum _{i}
 j_{ki}\left( {\bf r}\right),
\end{equation}
 where \( n_{k}({\bf r}) \) is the number of particles of species $k$ at node \( {\bf r} \),
 while $j_{ki}({\bf r})$ accounts for the fraction of particles of species $k$ going to node
 ${\bf r}+{\bf c}_i$.  If  we consider the velocity moving opposite to $i$, i.e.
 ${\bf c}_{i'}=-{\bf c}_i$, we have $j_{ki}({\bf r})=-j_{ki'}({\bf r}+{\bf c}_i)$ because
these fluxes are always defined considering that the particles move away from the reference node.
This unambiguously show that the fluxes are related to the links joining the connected nodes,
rather than being quantities defined on the nodes.

It is worth noting that in the previous balance equation the
relevant quantity is the number of particles of species $k$ at
node ${\bf r}$, $n_k({\bf r})$, rather than its number density,
$\rho_k({\bf r})$. If we take the volume of a cell as our unit of
volume, then $\rho_k({\bf r})=n_k({\bf r})$. However, in the
presence of solid boundaries this distinction may become relevant.
 The prefactor $A_{0}$ in Eq.~(\ref{discrete smolu}) is related to the geometrical structure
of the lattice. Rather than connecting it directly with the area
of the Wigner-Seitz cell that can be associated to node ${\bf r}$,
we derive its magnitude by computing how density diffuses to the
neighboring nodes. In Section~\ref{Sec: Effective diffusion} we
will compute explicitly this geometric prefactor for a particular
lattice. In the following, when referring to link mobility, we will use
the symbol $d_k=D_kA_0$.

Using link fluxes to compute the variation of the densities of the different species avoids approximating
 the divergence on a lattice, a source of lattice artifacts, and the related potential spurious fluxes
 that may appear.  Moreover, the use of these link fluxes also imposes locally mass conservation to
machine accuracy, avoiding the errors caused by the discretization
of the spatial gradient operator. We must still provide a
prescription to implement the diffusive fluxes.
 These are in principle given by Eq.~(\ref{flux})
and involve spatial gradients between two neighboring lattice
nodes. In equilibrium, $n_k^{eq}\sim\exp[-\beta\mu^{ex}]$ and, as
a consequence, Eq.~(\ref{flux}) predicts that all diffusive fluxes
vanish. However, the direct implementation of Eqn.~(\ref{flux}) on
a lattice will suffer from discretization errors that will result
in small but noticeable spurious fluxes. To eliminate this effect,
it is convenient to write the expression for the flux on a link as
\[
{\bf j}_{k}({\bf r},t)=-D_{k} e^{-\beta \mu ^{ex}_{k}({\bf r},t)}\nabla\left[ \rho_{k}({\bf r},t)e^{\beta \mu ^{ex}_{k}({\bf r},t)}\right].
\]
because, in this expression, the gradient becomes  identically
zero when the density distribution corresponds to its equilibrium
form. This also holds for the discretized form to be discussed
below.
 Consistent with the idea that the flux can be expressed in terms of link mass fluxes,
we propose a symmetrized implementation of $j_{ki}$ involving
magnitudes defined in the two connected nodes, ${\bf r}$ and ${\bf
r}+{\bf c}_i$. In particular, we write the flux of species
 $k$ along the link ${\bf c}_i$ as
\begin{equation}
\label{linkflux}
j_{ki}({\bf r})  =  -d_{k}\frac{e^{-\beta \mu_k^{ex}({\bf r})}+e^{-\beta \mu_k^{ex}({\bf r}+{\bf c}_i)}}{2}
                   \left[
                          \frac{n_k({\bf r}+{\bf c}_i) e^{\beta\mu_k^{ex}({\bf r}+{\bf c}_i)}-n_k({\bf r}) e^{\beta\mu_k^{ex}({\bf r})}}{\Delta_i}
                   \right],
\end{equation}
where  $\Delta_i=|{\bf c}_i|=|{\bf c}_{i'}|$ is the distance
between the two neighboring nodes. This symmetrized formulation
ensures that, to machine accuracy, $j_{ki}({\bf r})=-j_{ki'}({\bf
r}+{\bf c}_i)$, and mass is conserved for the model elementary
dynamic processes.
 Note that, based on the mass conservation expression, Eq.~(\ref{discrete smolu}), the global mass
change of node
 ${\bf r}$ is the sum of the link fluxes, $j_{ki}$. Mass evolution in the diffusive
limit is described only on the basis of mass flux divergence, as
we have described. In general, the procedure  developed based on
link fluxes provides a consistent framework to obtain other
gradients if needed.

\subsubsection{Advection.}

\begin{figure}
\includegraphics{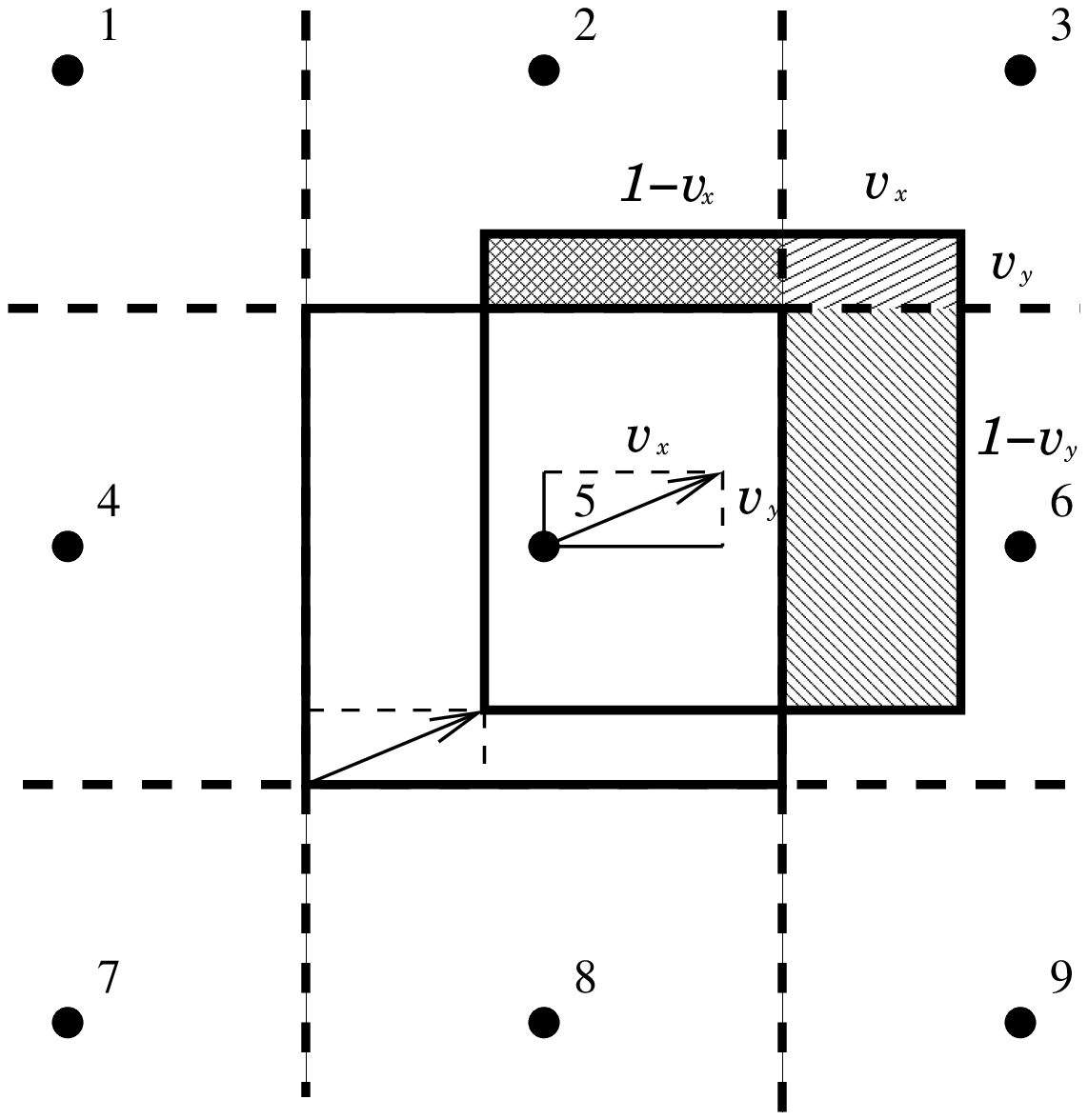}
\caption{Density redistribution due to advection.  To advect the
charge of a given node (in this case, node number 5) in one time
unit,  we shift the whole cell with the local velocity vector of
that node, $(v_x,v_y)$. Next, we displace a fraction of density
equal to the area of the cell that is now in the corresponding
site. In the graph a fraction of the density equal to the shadowed
rectangle area (\protect\( v_{x}v_{y}\protect \)) goes from cell 5
to cell 3, a fraction \protect\( (1-v_x)v_y\protect \) goes to
cell 2, \protect\( (1-v_y)v_x \protect \) goes to cell 6, and
\protect\( (1-v_x)(1-v_y)\protect \) stays at node 5. For the sake
of clarity, the figure shows a two-dimensional flow. In practice,
the analogous procedure is carried out in 3D.
 \label{fig1:advection}}
\end{figure}

Local density can also be altered due to advection if there is a
local velocity of the fluid. If,  for the time being, we disregard
diffusion, the advection mechanisms can be written in the form,
\begin{equation}
\label{advection}
\frac{\partial}{\partial t}\rho _{k}=-\nabla \cdot (\rho_{k}{\bf v}),
\end{equation}
where \( {\bf v} \) is the barycentric fluid velocity. In
principle, the change in the number of particles could be computed
on the basis of the advection along each link, in a way similar to
Eq.~(\ref{discrete smolu}). However, as we will describe in the
next section, the model we will introduce provides the velocity at
each node, rather than the link velocity. In order to avoid
numerical artifacts and spurious diffusion  due to the
interpolation to get such a link velocity, we propose an
alternative implementation of the advection process. We still
consider that $n_k({\bf r})$ give us the number of particles in a
volume element centered around node ${\bf r}$. Since we know the
velocity of that node, ${\bf v}({\bf r})$, in one step the node
will {\sl virtually} displace to ${\bf r}+{\bf v}({\bf r})$ . As a
result, the volume associated to node ${\bf r}$ will intersect
some neighboring cells of the real lattice (see
Fig.\ref{fig1:advection}). We then  distribute the amount of
particles $n_k$ into the intersected volumes proportionally to the
intersected region.  In Fig. \ref{fig1:advection}, we depict in
shadow the volumes that correspond to the fraction of the density
that is transported in the new cells. The advantage of this
approach is that it greatly reduces the spurious diffusion that
usually results during advection in lattice models.  To be more
precise, even with the present method, advection will cause some
spurious diffusion (proportional to  the flow velocity). However,
in Section~\ref{Sec: Effective diffusion} we show that, in
practice, this effect is negligible.

\subsection{Lattice Boltzmann Method.\label{LB paragraph}}
In order to simulate the hydrodynamic flow of the fluid, we make
use of the lattice-Boltzmann approach. This technique has been
used extensively to model hydrodynamic flows in complex geometries
~\cite{succi}. It is equivalent to solving a discretized version
of the Boltzmann equation with a linearized collision operator.
This method describes the dynamics of a fluid in terms of the
densities of particles that ``live'' on the nodes of a cubic
lattice and have discrete  velocities $\{{\bf c_i}\}$, where $i$
labels the links between a lattice point ${\bf r}$ and its
neighbors. The values of the velocities are chosen such that, in
one time step, a particle moves along a link from one lattice node
to its neighbor. In the Lattice-Boltzmann model, the unit of
length is equal to the lattice spacing and the unit of time is
equal to the time step. In addition, the unit of mass (or,
equivalently, energy) is fixed by the requirement that, in the
continuum limit, the transport equations for the lattice model
approach the Navier-Stokes equation. This imposes a relation
between the temperature and the speed of sound (see below
Eq.~\ref{N-S ideal}).
The central dynamic quantity in the Lattice-Boltzmann approach is
the one-particle distribution function,
 \( f_{i}({\bf r},t) \), which describes the probability
of having a particle at site \( {\bf r} \) at time $t$ with velocity
\( {\bf c}_{i} \). The hydrodynamic variables are obtained as moments of this
distribution function over the lattice velocities, ${\bf c}_i$; e.g. density and
momentum  can be obtained as
\begin{equation}
\label{rholb}
\rho ({\bf r},t)=\sum _{i}f_{i}({\bf r},t),  \;\;\;\;\;
{\bf j}({\bf r},t)\equiv\rho({\bf r},t){\bf v}({\bf r},t)=\sum _{i}{\bf c}_{i}f_{i}({\bf r},t),
\end{equation}
respectively.

In the presence of external forces, ${\bf F}$, the evolution equation can be expressed as
\begin{equation}
\label{lb1}
 f_{i}({\bf r}+{\bf c}_i,t+1)= f_{i}({\bf r},t)+{\cal L}_{ij}\left[ f_{j}({\bf r},t)-
f_{j}^{eq}({\bf r},t)\right]+\psi_i
\end{equation}
where ${\cal L}[\Psi]$ is a linear collision operator acting on $\Psi$ which tends to relax the distribution
function toward its equilibrium limit. Hence, one needs to specify the equilibrium
distribution as well as the collision operator. The collision operator ensures mass and momentum
conservation (i.e. $\sum_i{\cal L}_{ij}=\sum_i{\bf c}_i {\cal L}_{ij}=0$).
Its eigenvalues also determine the viscosity of the fluid. The equilibrium
distribution appearing in  Eq.~(\ref{lb1}) is that of an ideal gas. It can be shown that the
Navier-Stokes equations are recovered keeping a low velocity expansion of the Maxwellian~\cite{succi}, i.e.
\begin{equation}
f_i^{eq}=a^i\left[\rho+\frac{1}{c_s^2}{\bf c}_i\cdot{\bf j}+\frac{1}{2 c_s^4}\rho{\bf v}{\bf v}:({\bf c}_i{\bf c}_i-c_s^2{\bf 1})\right]
\end{equation}
where $:$ is the double inner product, and the coefficients $a^i$, depend on the geometry of the
lattice, and are chosen to ensure that the anisotropy of the
lattice does not affect the hydrodynamic behavior of the model, as
well as ensuring that all the distribution functions are non
negative. Moreover, $c_s$ is the speed of sound and its value
depends on the values of the coefficients $a^i$, but it is always
smaller than unity (in lattice units). Finally, the   term
$\psi_i$ accounts for the external force. It satisfies
$\sum_i\psi_i=0$ and $\sum_i{\bf c}_i\psi_i={\bf F}$. For a more
detailed description of how to model the external force, see
e.g.~\cite{LaddJSP,guo}.

It can be shown~\cite{LaddJSP} that in the hydrodynamic limit one recovers the Navier-Stokes equation
\begin{equation}
\label{N-S ideal}
\frac{\partial}{\partial t}\rho {\bf v}+\nabla\cdot\rho{\bf v}{\bf v}=\eta \nabla ^{2}\rho {\bf v}+\xi\nabla\nabla\cdot{\bf v}-c_{s}^{2}\nabla \rho +{\bf F}.
\end{equation}
Since the third term on the rhs is the pressure gradient for an
ideal gas, if we fix the temperature such that \( k_{B}T=c_{s}^{2}
\), we then recover Eq.~(\ref{General Navier-Stokes}) for an ideal
mixture. For non-ideal mixtures, we will introduce the missing
contribution to the pressure gradients as a local external force,
${\bf F}$.

Introducing the mixture non-ideality  as a  local effective force
implies that  the fluid reacts with the appropriate susceptibility
to applied external fields, although in the absence of spatial
gradients the equilibrium distribution corresponds to that of an
ideal gas. Since we are not concerned with local structure, the
model can be regarded as an effective kinetic model, similar in
structure to a linearized Vlasov equation. Hence, this approach
differs from previous proposals which try to derive the
hydrodynamics of non-ideal mixtures from kinetic models of
mixtures\cite{luo} or from a modification of the equilibrium
distribution to recover the equilibrium pressure\cite{swift}.

For a particular choice of the shear viscosity,
$\eta=1/6$ in lattice units~\cite{LaddJFMI}, the
 general  dynamic rule Eq.~(\ref{lb1}) simplifies to,
\begin{equation}
\label{eta16} f_{i}({\bf r},t+1)=a^{i}\left[\rho ({\bf r},t)
+\frac{1}{c_s^2}{\bf c}_{i}\cdot\left( {\bf j}({\bf r},t)+{\bf
F}\right)+\frac{1}{2c_s^4}\rho{\bf v}{\bf v}:({\bf c}_i{\bf
c}_i-c_s^2{\bf 1})\right].
\end{equation}
For the sake of convenience, we implement the model with this
simplified updating rule. However, it is straightforward to
implement the more general form that allows us to impose other
values of the viscosity.

The peculiarities of the non-ideality of the mixture enters
through the forcing term (${\bf F}$)in Eq.~(\ref{eta16}). This forcing term
can be decomposed into an external field and an interaction
contributions, ${\bf F}={\bf F}^{ext}+{\bf F}^{sol}$. This
interaction force, as previously described in
Eq.~(\ref{eq:gibbsduhem}), has the form ${\bf
F}^{sol}=\sum_k\rho_k\nabla\beta\mu^{ex}_k$. Using the same
approach that we have used to model the convection-diffusion
equation, we can determine the force acting on each link, $ F_i$.
Moreover, for the particular case where the diffusion matrix is
diagonal,
\begin{equation}
\label{eq:flink}
F_{i}({\bf r})=\sum_k\left[\frac{j_{ki}}{D_k}-\frac{n_k({\bf r}+{\bf c}_i)-n_k({\bf r})}{\Delta_i}\right].
\end{equation}
The advantage of using the force exerted on the links is that
again, we keep a symmetric dependence on the neighboring nodes,
and moreover $F_{i}({\bf r})=-F_{i'}({\bf r}+{\bf c}_i)$. Yet, in
the Lattice-Boltzmann update rule, we need the force acting on the
node. This force can be obtained averaging
the link forces,
\begin{equation}
F_{\alpha}^{sol}({\bf r})=\sum_i a^i c_{i\alpha} F_{i}({\bf r})\;\;\;\;\;\alpha=\mbox{x, y, z}
\end{equation}

Let us now introduce an alternative way of treating the same
systems. There are situations, as is the case in electrolytes,
where one of the components of the mixture is dominant, and plays
the role of the solvent. In this case, we can single out this
component, $\rho_s$, and treat it separately from the rest. In
particular, since $\rho_s\gg \rho_k$, we can approximate the
overall density by the solvent density ($\rho \simeq \rho_s $ ),
and the overall momentum by the solvent momentum ($\rho v = \sum_k
\rho_k v_k \simeq \rho_s v_s$). If we then relate the moments of
the distribution function $f_i$ to the solvent density, i.e.
\(\sum_if_i=\rho_s\) and \(\sum_i{\bf c}_if_i=\rho_s{\bf v}_s\)
instead of Eqs.~(\ref{rholb}), we impose a constant solvent
density in the incompressible regime. Hence, the rest of the
components will need to compensate their densities to avoid any
net local density variation. Although this incompressibility
constraint is not exact, it may be a convenient approximation.
From the point of view of the link force, Eq.~(\ref{eq:flink}), it
has the computational advantage that one gets
$F_{i}=\sum_kj_{ki}/D_k$ and it reduces  to the link diffusive
flux previously computed, Eq.~(\ref{linkflux}). In this case the
Navier-Stokes equation becomes
\begin{equation}
\frac{\partial}{\partial t}\rho_s{\bf v}_s+\nabla\cdot\rho_s{\bf v}_s{\bf v}_s=\eta\nabla^2{\bf v}_s+
\xi\nabla\nabla\cdot{\bf v}_s-c_s^2\nabla\rho_s-k_BT\sum_k\left[\nabla\rho_k+\nabla\mu_k^{ex}\right]+{\bf F}^{ext},
\end{equation}
and by taking $k_BT=c_s^2$, we recover an appropriate behavior when $\rho_s\gg\rho_k$.

The advantage of this approach is that densities of different
species are dealt with on different footing, which may prove
advantageous in certain applications, specially when dealing with
boundary conditions that act differently on the solvent and
solute, as it is the case if dealing with semipermeable membranes.
Numerically, in this case there is a net force only when the
density distribution deviates from its local equilibrium value, in
contrast with the original method, where the density coming from
the advection contribution balances the local force. This ensures
an additional way to avoid spurious artifacts from the underlying
lattice.

\section{Boundary conditions}

If the fluid mixture is confined between walls, or if colloids are added to the mixture,
we need to specify how the densities and distribution function will interact with solid
interfaces. To account fully for such an interaction, we need to describe in turn how the
distribution function behaves, how the particle number evolves, and how we estimate
the interacting force at the surface.

At a solid surface we expect hydrodynamic ``stick'' boundary
conditions to apply. One way to impose these is to  apply the
so-called ``bounce-back rule'' on the links. However, the standard
version of this procedure (see for example Ladd~\cite{LaddJSP})
allows the fluid to leak into the solid. Although this leakage is
usually innocuous, there are cases (a typical example being when
electrostatics is part of the excess chemical potential) where
this leakage may change the density of the solvent inside the
solid, leading to a  corresponding error in the pressure gradient.
There exist alternative bounce-back rules that do not allow for
any fluid leakage~\cite{ngai}.

The formulation of our model in terms of link fluxes simplifies
the implementation of boundary conditions for the fluxes of the
different species densities, $\rho_k$. Since the
convection-diffusion equation involves only mass conservation, it
is enough to impose that there is no net flux on any link that
joins a fluid node and a solid node. We accomplish this by
imposing that the diffusive flux $j_{ki}=0$ on such a link, and
that the flux due to advection also vanishes. This second
requirement is achieved by a kind of partial bounce-back move: the
number of particles that would have been assigned to a solid node
after advection is reflected back to its node of origin.

The updating rule, both for the number densities of the
convection-diffusion equations and for the lattice Boltzmann
distribution function, requires the evaluation of gradients of
chemical potentials. To this end, we need to specify the values of
the excess chemical potentials on neighboring nodes, and those may
involve the values of the fluid densities in contact with the
solid wall. We consider that the relevant value of the density is
that in contact with the wall, which is somewhere in between the
fluid and the solid node. Such value can be obtained by requiring
that it is consistent with the no-flux condition for the link flux
of that species. The no flux condition is satisfied requiring (see
Eq.~(\ref{linkflux})),
\begin{equation}
n_k({\bf r}+{\bf c}_i)=n_k({\bf r})e^{\beta\left[\mu_k^{ex}({\bf r}+{\bf c}_i)-\mu_k^{ex}({\bf r})\right]},
\end{equation}
which should be understood as the extrapolation of the fluid
density to ensure the absence of flux diffusion, and in general it
is an implicit equation to obtain an estimate  of the extrapolated
number of particles, $n_k({\bf r}+{\bf c}_i)$. Note that this
fictitious extrapolated density is a property of the link, not of
the node.

As we have mentioned in Section \ref{section:model}, the
formulation based on the fluxes is based on the evolution of the
number of particles contained in a given volume element. For the
fluid nodes in the absence of solid interfaces the particle number
is proportional to the number density. This is no longer the case
close to a solid wall. This difference is pertinent because the
excess chemical potential and the pressure are functions of the
number density, $\rho_k$. While for a wall at rest, one can still
consider that the wall is equidistant from the nodes and $n_k$ and
$\rho_k$ coincide, for a moving solid surface, the position of the
solid boundary will change as it moves. In this case,  a
coefficient $\alpha$ that establishes how close the solid boundary
is to the fluid node should be introduced. In the limiting case
that the solid boundary is reaching the neighboring fluid node,
the corresponding cell has a volume that is approximately half the volume of a usual cell,
hence  $\alpha=1/2$;
in the opposite case when the solid surface reaches the solid node one gets accordingly $\alpha=3/2$.
This coefficient then allows us to relate
$n_k=\alpha \rho_k$. Although there exist different ways in which
this coefficient may be computed, any smooth function that
accounts for the volume change will be enough to avoid  abrupt
changes in the density when a fluid nodes is absorbed or created
by the moving boundary.

\section{Electrokinetic Equations.}

In the previous sections we have developed a model to simulate
general non-ideal fluid mixtures. We will now analyze the special
case in which the fluid mixture is an electrolyte. The simplest
electrolyte model corresponds to a three-species mixture, two of
them being  the ionic species, $\rho_+$ and $\rho_-$ with charges
$z_+e$ and $z_-e$, and the third one being the neutral solvent
$\rho_s$. $e$ is the elementary charge, and $z_+$ and $z_-$ are
the valencies of the ions.  The local charge can then be expressed
as $q({\bf r})=e[z_+\rho_+({\bf r})+z_-\rho_-({\bf r})]$. The
simplest free-energy model corresponds to an ideal mixture in the
absence of  any local electric field, where we can write,
\begin{equation}
\label{eq:freeenergypb}
\beta f({\bf r})=\sum_{k=\pm,s}\rho_k[\log(\Lambda_k^3\rho_k)-1]+\frac{1}{2}\beta q\widehat{\Phi}
\end{equation}
with $\widehat{\Phi}$ being the electrostatic potential and the factor $1/2$ avoids double counting.
The chemical potential is then
\begin{eqnarray}
\beta \mu_k&=&\log\rho_k+\beta z_k \widehat{\Phi},\;\;\;k=+,-,\;\;\;\;\;\; \beta \mu_s=\log\rho_s\nonumber\\
\end{eqnarray}
The hydrodynamic evolution equations for this free energy model become,
\begin{eqnarray}
\label{electrokin}
\frac{\partial}{\partial t}\rho_k+\nabla\cdot\rho_k{\bf v}&=&D_k\nabla\cdot\left[\nabla\rho_k+
ez_k\rho_k\nabla\beta\widehat{\Phi}\right]\\
\frac{\partial}{\partial t}\rho{\bf v}+\nabla\cdot\rho{\bf v}{\bf v}&=&\eta\nabla^2\rho{\bf v}+
\xi\nabla\nabla\cdot\rho{\bf v}-c_s^2\nabla\rho+\beta\sum_kez_k\rho_k\nabla\widehat{\Phi}.
\label{electrokinvel}
\end{eqnarray}

We still need an additional equation that prescribes how the
electrostatic potential is related to the local charge density.
Since transport processes associated to mass and momentum transfer
in fluid mixtures are much slower than  the propagation of
electromagnetic waves, the electric field is completely determined
by the Poisson equation
\begin{equation} \label{poisson}
\nabla^2\Phi=-4\pi l_B\left[\sum_{k=\pm}z_k\rho_k+\rho_s\right]
\end{equation}
which has been expressed in terms of a dimensionless potential, $\Phi=e\beta\widehat{\Phi}$,
while  $l_B=\beta e^2/(4\pi\epsilon)$ is the Bjerrum length (the distance at which the electrostatic and
the thermal energies are equal), with $\epsilon$ the dielectric
constant of the fluid.
In the previous equation, $\rho_s$
stands for the charge density of the solid surfaces, if there are
confining walls or moving suspended particles in the electrolyte.
Obviously,
 $\sigma$ will be non-zero only on those solid surfaces.
The Equations (\ref{eq:freeenergypb}), (\ref{electrokin}),  and
(\ref{poisson}) are commonly referred to as the Electrokinetic
equations.

The electrostatic potential \( \Phi  \) can be computed using
standard techniques. Specifically, we have implemented a
successive over-relaxation scheme (SOR)\cite{Numerical Recipes} as
described in more detail in Ref.~\cite{Juergen}. The advantage of
this model is that it does not presume a specific type of boundary
condition, and can be easily generalized to deal with media of
different dielectric constants. Although not as fast as other
methods for solving the Poisson equation, it is adequate for our
purposes because, once the local equilibrium charge profiles are
achieved, the calculation of the disturbed electrostatic potential
due to external forces is much less time consuming than the
iteration part related to lattice-Boltzmann and
convection-diffusion.

\section{Validation tests}

In order to validate the model that we introduced in the previous
section, we compare its predictions against known results. In
particular, we verify that the equilibrium charge distribution is
properly recovered on the lattice, and that out of equilibrium the
different coupling mechanisms between fluid flow and charge
inhomogeneities are properly accounted for.

\subsection{Effective diffusion} \label{Sec: Effective diffusion}

As was pointed out below Eq.~(\ref{discrete smolu}), the
diffusion coefficient characterizing the discrete version of the
diffusion equation is not the same as the link diffusion
coefficient, $d_k$ but is related to it through a simple
geometrical factor $A_0$ that depends on the type lattice used.
$A_0$ can be evaluated as follows. Consider a situation where the
transport of species $k$  is purely diffusive.  A density
perturbation $\rho_0$, initially localized at node ${\bf r}_0$, will
spread in one time step to the connected neighboring nodes. If the
process is purely diffusive, we know the amplitude of the second
moment of the density variation during this time step and $\sum_i
\Delta_i^2 \rho({\bf r}_0+{\bf
c}_i,t_0+1)=6D_k\rho_0=6A_0d_k\rho_0$ in a three dimensional cubic
lattice. Let us consider for concreteness the D3Q18
lattice~\cite{LaddJFMII}, which is the lattice we used in our LB
simulation. Since the link fluxes
 $j_{i}=d_k\rho_0/\Delta_i$, after one time step the density in each of the six
nearest neighbors is $d_k\rho_0$, while the density in each of the
other 12 connected nodes is $d_k\rho_0/\sqrt{2}$. As a result,
$\sum_i \Delta_i^2 \rho({\bf r}_0+{\bf c}_i,t_0+1)=d_k
(6+12\sqrt{2})\rho_0$, which implies that $A_0=1+2\sqrt{2}$
(or $D_k=d_k(1+2\sqrt(2)$).
Depending on  the value of $d_k$ it might happen that the total
density transferred to the neighbors is larger than the initial
density. For D3Q18 this gives us an upper bound for the input
diffusion coefficient that ensures absolute stability, $d_k\leq
1/(6(1+2\sqrt{2}))=0.044$. In practice, we find that for all cases
that we have analyzed, numerical instabilities related to
diffusion become relevant for values of the input diffusion
coefficient $d_k\geq 0.05$. In order to perform simulations at
higher diffusivities, we need to modify the numerical scheme to
simulate the diffusion equation. This instability can be overcome
by introducing a multiple-timestep technique. To this end, we
introduce a smaller diffusion coefficient \( d_{it}=d_k/N_{it} \)
and iterate \( N_{it} \) times the discrete diffusion equation,
Eq.~(\ref{discrete smolu}), to advance the densities one time
step.

When applying this multiple timestep method to solve the lattice
diffusion equation, one must compute carefully the force that
should be applied to the distribution function
 $f_i$ at the end of the time step. In fact, ${\bf F}^{sol}$ should be computed at all
the intermediate steps.  All these contributions should then be
added to obtain the total force at the end of the iteration. With
this technique can vary the diffusion coefficient over several
orders of magnitude.  For example, in our simulations we could
vary $D_k$ from $D_k= 10^{-3}$ to $D_k=6$ (all in lattice units).

On top of the lattice effects on diffusion itself, advection can
also induce spurious diffusion, because  the lattice velocities do
not coincide in general with the local velocity. As a consequence,
a concentrated set of particles will spread over the lattice
nodes, even if subject to a pure translational motion. Hence, only
when the velocity is commensurate with the lattice spacing, both
in direction and magnitude, will spurious diffusion be exactly
zero. We must then quantify the amount of spurious diffusion. To
this end, we consider an ideal binary mixture composed of a
solvent with initial uniform density, $\rho_s$, and a solute with
initial density $\rho_t$. The mixture is contained  between two
parallel walls that are permeable to the solvent but impermeable
to the solute. The fluid is moving with a uniform
velocity $v$ perpendicular to the walls. As a result of the
impermeability of the walls to the solute, a steady
state is reached, determined by the solvent density profile,
$\rho_t(x)$, which satisfies
\begin{equation} \label{Equilibrium effective}
\rho_{t} (x) = \rho_{0} \exp \left[-\frac{v}{D^*}(x-x_0) \right],
\end{equation}
where $v$ is the fluid velocity, $D^*$ the effective diffusion coefficient, $\rho_0$ the
solvent distribution at contact with the wall located at $x_0$.

\begin{figure}

\includegraphics[width=0.8\textwidth]{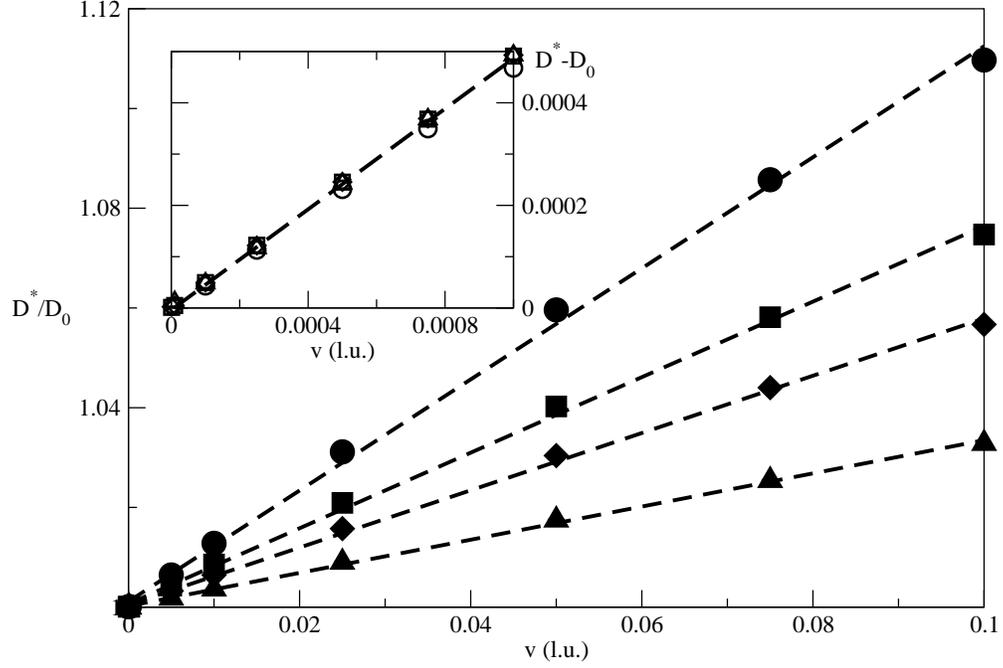}

\caption{In the Lattice-Boltzmann model, advection causes some spurious diffusion.
The figure shows the
computed effective diffusion, $D^*/D_0$ as function of the fluid
velocity for the steady state described in Section~\ref{Sec:
Diffusion}. The curves are drawn for different diffusion
coefficients at zero velocity: $D_0=0.38$ (circles), $D_0=0.57$
(squares), $D_0=0.76$ (diamonds), and $D_0=1.34$ (triangles).  In
the inset we show that the amount of diffusion induced by the flow
does not depend on the equilibrium coefficient and has, for small velocities, a linear
velocity dependence. Symbols are simulation result and the dashed
line corresponds to the theoretical expression
described in the text.
\label{fig:Effective diff} }

\end{figure}

In Fig.~\ref{fig:Effective diff} we show the effective diffusion
coefficient measured by using Eq.~(\ref{Equilibrium effective}) as
function of the fluid velocity for a range of values of the
diffusion coefficient. We plot $D^*/D_0$ (where $D_0$ is the
diffusion coefficient for a quiescent fluid). In order to show
that there exists an intrinsic advection-induced spurious
diffusion, we plot in the inset of the same figure
 the difference between the effective and the input diffusion
coefficient for many values of the input diffusion coefficient as
function of the fluid velocity. Because all curves collapses, this
graph shows that the diffusion coefficient induced by the
advection depends only on the fluid velocity. We observe that the
dependence on the (absolute value of) flow velocity is linear with slope $1/2$.
Following the procedure that we used above to compute the factor
$A_0$, we can derive an expression for the advection-induced
diffusion coefficient. In one dimension, a fraction $v\Delta t$ of
the density $\rho(x)$ is displaced to the next node, while a
fraction $(1-v)\Delta t$ remains at the original node. The center
of mass of the density is displaced by a factor $v\Delta t$.
Simple algebra then shows that the second moment of the density
variation during a time step is $<\Delta_i^2>=v(1-v)$. The
flow-induced diffusion coefficient in one dimension is therefore
$D^*=1/2v-1/2v^2$. In three dimensions this expression is readily
generalized to yield
\begin{equation}
D^*= \frac{1}{2} \left[ v_x(1-v_x) + v_y(1-v_y) + v_z(1-v_z) \right].
\end{equation}
By choosing a sufficiently low value of the flow
velocity, and a sufficiently large value of $D_0$, we can largely
suppress the effect of this advective diffusion.

\subsection{Electrolyte in a slit.\label{Slit paragraph}}

Next, we consider a fluid confined between two parallel solid
walls at rest, with a constant surface charge.  The slit has a
width $L$ and the surface density charge  is fixed to \( \rho
(-L/2)=\rho (L/2)=\sigma /2 \).

The space between the two slits is occupied by a solvent and counter-ions.
In order to achieve global neutrality, the density of counter-ion
is initially set to be uniformly distributed \( \rho (x)=-\sigma /L \), \( x\in
\{-L/2,L/2\} \).

The actual position of the hydrodynamic and electrostatic solid
boundary cannot be resolved within a lattice spacing.
In the neutral case, for the
viscosity and geometry considered the wall can be assumed to be
halfway between two consecutive lattice nodes, as dictated by the
bounce-back rule\cite{LaddJFMI}. We will use this position as a
reasonable approximation. In fact, the results we describe for a
planar slit indicate that for a planar wall the electrostatic
position of the wall can be taken as being midway between the
boundary nodes. For a non-planar interface a separate calibration
will be required.

\subsubsection{Equilibrium distribution of the counter-ion density}
In equilibrium,  a uniform charge density on a flat  wall will
induce an inhomogeneous equilibrium density profile of the
counter-ions. For this simple geometry,  the charge-density
profile of the counter ions is known analytically known (at least,
at the Poisson-Boltzmann level) \cite{Israelachvili, PBWarren} for
an arbitrary surface charge density:
\begin{equation}
\label{eq:rhoprofileslit}
\rho (x)=\frac{\rho _{0}}{\cos^{2}(Kx)}
\end{equation}
where \( \rho _{0}=K^{2}/2\pi l_{B} \), $K$ is the solution of the
transcendental equation
\begin{equation} \label{KL solution}
\frac{KL}{2} \tan\left( \frac{KL}{2}\right) =\pi l_{B}L\sigma
\label{eqK}
\end{equation}
which involves the wall charge density. Since we have an exact
solution for the full Poisson-Boltzmann equation
for arbitrary values of the wall charge, this geometry is a
good case to analyze the limitations of the model dealing with
large charges, i.e. beyond the linear Poisson-Boltzmann limit.
For low surface charge densities, the linear regime is
recovered by linearizing Eq.~\ref{KL solution}, and the parameter $K$ becomes $K_{lin}L=\sqrt{4\pi
l_B\sigma}$.

In the opposite limit of high  surface-charge density, $K$
saturates at $K_{sat}L=\pi$.  We can then quantify the deviation
of the fluid from the linearized regime where the electrostatic
interactions are small by analyzing the departure of $KL$ from
$K_{lin}L$.

\begin{figure}

\includegraphics[width=0.9\textwidth]{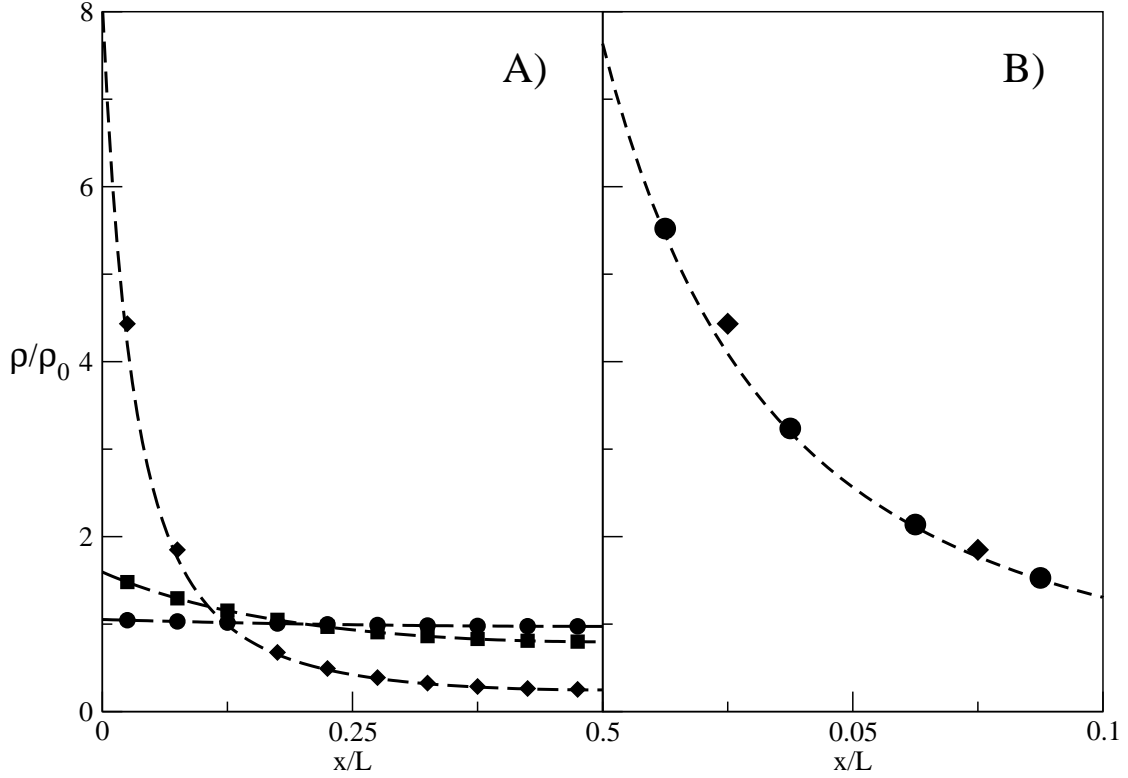}

\caption{Equilibrium distribution of the charge density of
counter-ions(no added salt) in slit between two charged walls at a
distance $L$. The abscissa measures the distance from the wall in
units of $L$. The local density is expressed in units of the
average charge density in the bulk:  \protect\( \rho _{0}=\sigma
/L\protect \). A) charge distributions  for three values of the
dimensionless parameter $KL$ (see text):  $KL=0.553$ (circles),
$KL=1.57$ (squares) and $KL=2.77$ (diamonds). In the same figure,
we have indicated the corresponding analytical results (Eq.~\ref{eq:rhoprofileslit}) (dashed curves)
for a slit of width $L=20$ lattice spacings. Circles and squares
correspond to the linear regime ($K/K_{lin}=1.01$ and $1.13$
respectively), while diamonds are close to the saturation limit
($K/K_{lin}=2.01$).  B) The accuracy of the numerical solution for
the charge profile can be improved  by increasing the spatial resolution of the
lattice, in this case from L=20 (diamonds) to L=40 (circles).
Again, the analytical result is shown as a dashed curve.  The curves in
B correspond to the result for  a highly charged surface,
$KL=2.77$ ($K/K_{lin}=2.01$).
\label{fig2:equilibrium density distrib} }
\end{figure}

In Fig.~\ref{fig2:equilibrium density distrib}.A we show the
equilibrium counter-ions distributions in both limits.  In our
simulations we fixed the Bjerrum length to be 0.4, the channel
width to 20 lattice nodes, and we have varied the surface-charge
density. In the plot we show the profiles for $K/K_{lin}=1.01,
1.13$ and $2.01$, which correspond to $\sigma=0.003125,$ $
0.03125$ and $0.3125$ in dimensionless lattice units, respectively.  The highest value of $K$ is
not far from the saturation value. The figure shows that, with the
present method, we can indeed reproduce the correct counter-ion
distribution, both in the linear and in the non-linear regime. In
Fig.~\ref{fig2:equilibrium density distrib}.B we compare the
density profiles close to the wall in the non-linear regime for
two different slit widths. The larger the surface charge the more
localized the charge profile will be. The figure shows that
increasing the resolution of the lattice does result in a small
but significant improvement in the calculation of the charge
distribution.
 Of course, the
discrepancy would be greater for a more localized charge profile.
In practice, only the computer resources (memory) will set an
upper limit for the surface charge density that can be modeled
reliably with the present scheme.

\subsubsection{Electro-osmotic flow}

Having verified that the model
correctly reproduces the equilibrium behavior,
we next turn to the calculation of flow caused by an external
electric field. We apply a constant external electric field that
is parallel to the slit, \( E^{||} \). This field causes
hydrodynamic flow as it exerts a force on those fluid elements
that carry a net charge.  If we take $y$ as the component along
the walls and refer to $x$ as the coordinate perpendicular to the
walls, then, at the Poisson-Boltzmann level,  the exact solution
for the fluid flow in the steady state can be written as \cite{PBWarren}:
\begin{equation} \label{v flow} v_y(x)=\frac{eE^{||}\rho
_{0}}{\eta K^{2}}\log\left[
\frac{\cos(Kx)}{\cos\left(\frac{KL}{2}\right)}\right] ,
\end{equation} where \( \eta  \) is the shear viscosity of the fluid.
In our simulations, we model the constant electric
field by taking into account the potential difference that it
causes between neighboring lattice nodes (i.e.
$\Delta\widehat{\Phi}_{ext}(y)=E^{||}\Delta y$).

\begin{figure}

\includegraphics[width=0.9\textwidth]{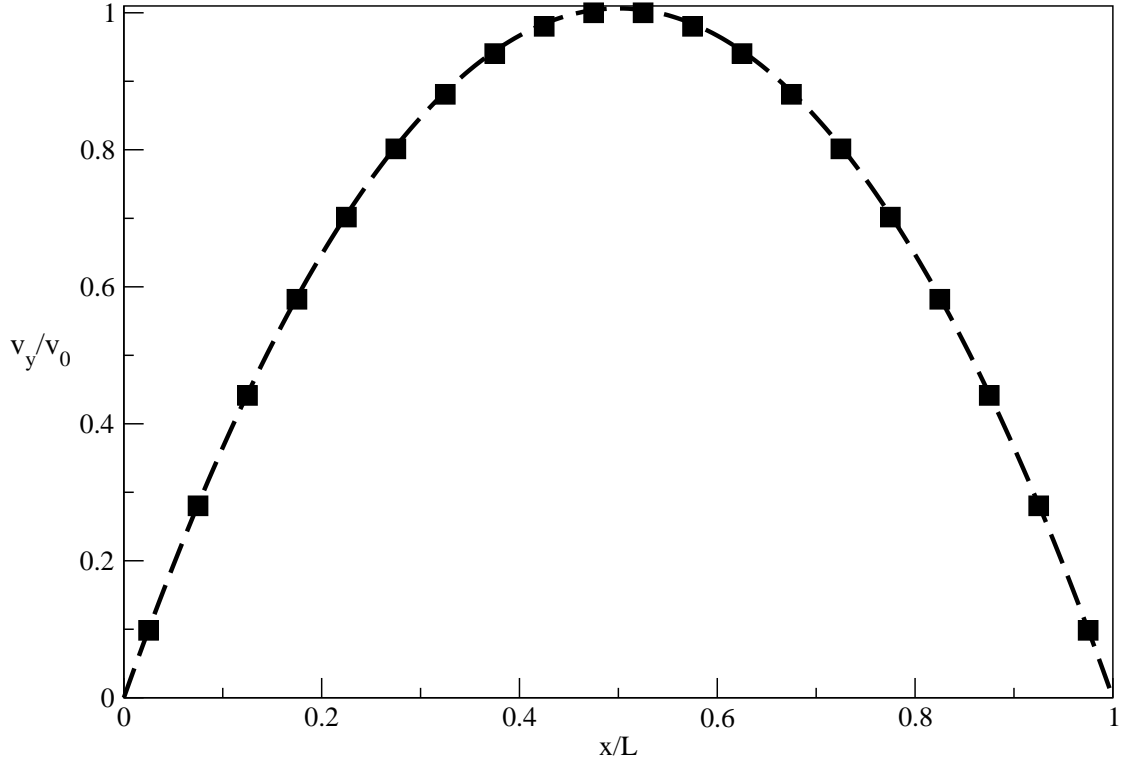}

\caption{Electro-osmotic flow profile in a slit of width $L=20$ lattice spacings.
The surface charge density, $\sigma=0.003125$ ($K/K_{lin}=1.01$),
corresponds to the linear regime. The fluid in between the slit
contains only counter ions. The electric field is along the
$y$-direction. It has a strength of 0.1 in units $k_BT/(\Delta l e)$,
where $\Delta l$ is the lattice spacing and $e$ is the elementary charge.
The simulation results are compared to the theoretical prediction,
Eq.(~\ref{v flow}), shown as a dashed curve.
\label{fig3:electrosmosisflow}}
\end{figure}

Fig.~\ref{fig3:electrosmosisflow}  shows the computed
electro-osmotic flow profile in a slit confined by hard walls with
a charge density $\sigma=0.003125$ (in units of the elementary
charge per square lattice unit). In the same figure, we also show
the analytical solution (Eqn.~(\ref{v flow}), with
$K/K_{lin}=1.01$) that is exact in the Debye-H\"{u}ckel limit.
Again, there is  good agreement between theory and simulation.
This suggests that the effect of electrostatic forces on the
hydrodynamic flow is correctly taken into account in the
simulations.

\subsection{Sedimentation velocity}

In the previous sections we have seen that the appropriate
equilibrium charge distribution is reproduced both in the linear
and non-linear regimes of the Poisson-Boltzmann equation, and that
also a charge distribution induces the correct fluid profiles. We
must still show that the opposite coupling works correctly, i.e.
we must compute the hydrodynamic drag on a charged object in the
absence of external electrical fields.

To this end, we compute the sedimentation velocity of an array of
charged spheres immersed in an electrolyte solution. In this case,
the velocity of the colloidal particle induces a fluid flow that
determines the steady charge distribution around the sphere. This
charge distribution in turn affects the sedimentation velocity of
the particle. Hence, all the different couplings between charge,
electrostatic potential and fluid flow are present. Such a
scenario has been analyzed previously with a different
model \cite{Juergen} and analytically at infinite
dilution \cite{Booth}. As a consequence, we can again check our
simulations against known results.

The system that we consider consists of a charged sphere of radius
\( a \) in a three-dimensional box of size \( L \). Because of
periodic boundary conditions, this corresponds to a periodic array
of spheres with volume fraction \( \varphi =(4\pi a^{3}/L^{3}) \).
In the simulation, we first allow the electrolyte to equilibrate
with the particle at rest in the absence of external forces; hence
the system develops its equilibrium double layer. Then, we apply
the gravity as an external body force applied to the fluid, i.e.
we move in the system of reference of the colloid. In this way we
avoid the problem of updating the particle's position due to its
motion \cite{LaddJFMII}. By forcing the colloid to be at rest, we
will not conserve momentum, but by computing the mean fluid
velocity in the steady state (which is reached on a timescale of
order $L^2\rho/\eta$), we can obtain the sedimentation velocity.

We have fixed the Bjerrum length to \( l_{B}=0.4 \) and the radius
of the sphere to \( a = 4.5 \) in lattice units.
We performed calculations for two different values of
the solvent fluid density, \( \rho _{s}=1 \), and \( \rho _{s}=20
\), while the density of the added salt $\rho_k$
was varied between $1.8*10^{-2}$ and $4*10^{-4}$. As we vary the salt
concentration, we also change the Debye length from $3.3$ to $21$.
Since $\rho_s\gg \rho_k$, we have performed most
calculations using the second version of our simulation scheme, as
described at the end of Section~\ref{LB paragraph}. However, we
also performed some simulations using the original model (taking
the solvent density as the overall density). The only difference
that we observe between the two implementations is a small
variation in the numerical value of the sedimentation velocity.
However, this difference already shows up for sedimentation of a
neutral sphere. It is due to a small change in the fluid viscosity
that is caused by a small difference in the overall fluid density
in the two implementations. The valency of the macro-ion was
chosen to be Z=10 which corresponds to the small charge
limit. Although our computational scheme should also work outside
the Debye-H\"{u}ckel limit, we restrict ourselves to this regime,
because it is only in this limit that we can compare with existing
analytical results. Specifically, Booth predicted that the
sedimentation velocity, $U_{0}(Z)$, of a weakly charged sphere of
valency $Z$ in the dilute limit can be expressed as~\cite{Booth}
\begin{equation} \label{eq:booth}
\frac{U_{0}(Z)}{U_{0}}=1-c_{2}Z^{2},
\end{equation} where \( U_{0} \) is the sedimentation velocity of
a neutral sphere, and \( c_{2} \) is a constant that can be
computed analytically in the Debye H\"{u}ckel limit. For the
simplified situation of monovalent co- and counter-ions,
$z_+=-z_-=1$, which have the same diffusivity, \( D_+=D_-=D \),
the expression for \( c_{2} \) simplifies to
\begin{equation} \label{booth simple}
c_{2}=\frac{k_{B}Tl_{B}}{72\pi a^{2}\eta D}f(\kappa a).
\end{equation} where \( f(\kappa a) \) is a linear combination of
exponential integral functions~\cite{Juergen} and is a function of
the inverse Debye length, \( \kappa =\lambda ^{-1}_{D}=\sqrt{4\pi
l_B\sum_kz_k^2\rho_k} \).  We have checked that the
sedimentation velocity scales as predicted with the viscosity. We
have also verified that we are indeed in the linear regime where
the sedimentation velocity is proportional to the applied
gravitational field. In particular, for the two values of the
density considered, $\rho_s$, the linear regime was obtained for
forces per unit of volume such that the flow velocity never exceeded
$0.1$ in lattice units.

\begin{figure}
\includegraphics[width=0.9\textwidth]{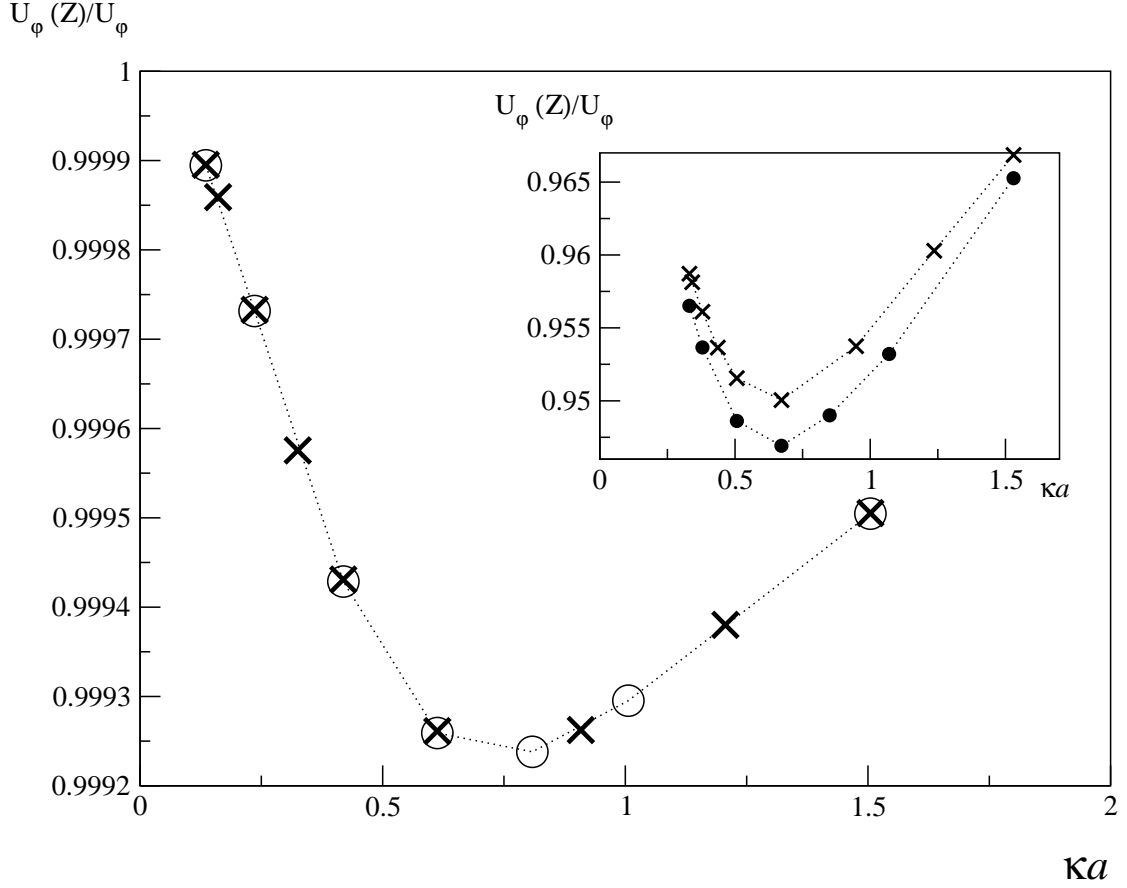}
\caption{Reduced sedimentation
velocity of a periodic array of colloids of valency $Z=10$ in an electrolyte  as a
function of \( \kappa a\protect \). The figure shows the results
for two different values of the ionic diffusion coefficients. The
curve for $D_0^{(1)}=0.95$ (circles) has been rescaled to the
curve for $D_0=0.19$ (x) according to Eq.~(\ref{booth simple}),
i.e. \protect\( U(D)=U(D_0^{(1)})*\left( D_0^{(1)}/D_0 \right)\protect\).
The superposition of the two
curves shows that the scaling in obeyed. In the inset we also
show the results for a colloid of valency $Z=100$. However, in
this high-charge regime the sedimentation velocity does not scale
with the diffusion coefficient in the way predicted by the
linearized theory.
\label{fig4:diffusion dependence} }
\end{figure}

Fig.~\ref{fig4:diffusion dependence} shows the sedimentation
velocity of a weakly charged sphere ($Z=10$) as a function of the
inverse Debye screening length.  As can be seen from the figure,
the sedimentation velocities scales with the ionic diffusivity in
the way predicted by Eq.~\ref{booth simple}. The inset in the same
figure shows that this scaling breaks down at higher colloidal charges (
$Z=100$), i.e. outside the range of validity of the linearized
Poisson-Boltzmann description.

Fig.~\ref{fig5:booth test} shows  the reduced sedimentation
velocity (\( U_{\varphi }(Z)/U_{\varphi }(Z=0) \)) as a function
of $\kappa a$  for a range of volume fractions. As the volume
fraction decreases, the curves approach Booth's infinite-dilution
result, while the minimum sedimentation velocity  moves toward the
minimum value predicted by theory. In order to compare
quantitatively the simulation results with Booth's theory,
Eq.~(\ref{eq:booth}), we must extrapolate the computed values for
$U_{\varphi}(Z)/U_{\varphi}(Z=0)$ from the finite $\varphi$-values
of the simulations to the infinite-dilution limit,
$U_{0}(Z)/U_{0}(Z=0)$. For neutral spheres Hashimoto has shown
that that the sedimentation velocity converges very slowly to its
infinite-dilution value, namely as~\cite{Hashimoto}
\begin{equation}
\label{U in the dilute limit}
\frac{U_{\varphi}(Z=0)}{U_{0}(Z=0)}=1-1.7601\varphi ^{1/3}+\varphi +O(\varphi
^{2}),
\end{equation}
Ladd has numerically verified this dependence~\cite{LaddJCP}. For
charged spheres, due to the electrostatic screening,
 we still expect that the dominant $\varphi$ dependence
comes from excluded volume; previous results indicate that this is
indeed the case \cite{Juergen}. When performing the dilute limit
expansion, we therefore decided to single out the major volume
fraction dependence by normalizing the simulation results with the
Stokes drag coefficient, i.e. computing the low density limit of
$U_{\varphi}(Z)/U_{0}(0)$. As a result, it is reasonable to obtain
the same functional dependence on $\varphi$ as Hashimoto with a
slightly different amplitude.     Specifically, we expect
\begin{equation}
\label{U charged dilute limit}
\frac{U_{\varphi}(Z)}{U_{0}(0)}=1-(1.7601+\epsilon)\varphi ^{1/3}+O(\varphi^{2/3}),
\end{equation}
where $\epsilon$ is much less than one. Eventually, the dilute
limit is obtained by extrapolating Eq.~\ref{U charged dilute
limit} to $\varphi=0$.

In Fig.~\ref{fig5:booth test} we show the extrapolated
sedimentation velocities for a particular value of $\kappa a$. The
estimated error in the limiting sedimentation velocity  is rather
large. It could have been reduced by computing more values of the
sedimentation velocity at low volume fractions. In addition, there
is some uncertainty in the value of the effective sphere radius.
In the light of these uncertainties,  the agreement with the Booth
limit in Fig.~\ref{fig5:booth test} is gratifying.

\subsection{Absence of spurious fluxes}

We pointed out in Section~\ref{section:model} that one of the
incentives for developing the present model was to eliminate any
mixing of continuous-space gradients and discretized gradient
operators. The reason is that the inevitable approximations
associated with the discretization of gradient operators usually
lead to the appearance of spurious  mass and momentum fluxes, even
in equilibrium. Such spurious fluxes are present in particular
whenever there exist spatial inhomogeneities related for example
to the presence of liquid interfaces. In the present approach, we
only use lattice-gradient operators that have been constructed
such that, in equilibrium, no flow can result. To demonstrate the
effect that this has, we compare the present method with an
existing ``mixed'' method. In particular,  we consider  a
spherical colloid of radius $a=4.5$, at rest in an electrolyte in
a cubic box of diameter $L=20$. The valency of the sphere is
$Z=10$ and the system as a whole is electrically neutral. In
Fig.~\ref{fig6:spurious currents} we show the projection of the
momentum flux in the equatorial plane of the sphere  and compare
these residual fluxes both for the model introduced in this paper
and the model of Ref.~\cite{Juergen}.
Fig.~\ref{fig6:spurious currents}.a shows that spurious currents,
although
small, are certainly not negligible in this case. Moreover,  their
magnitude is clearly correlated with the distance to the colloidal
particle: the largest currents appear in the region where the
spatial gradients are largest. For highly charged spheres (i.e.
outside the linear Debye-H\"{u}ckel regime) these spurious fluxes
will become larger.  In contrast, in Fig.~\ref{fig6:spurious
currents}.b (present model), the spurious fluxes are at the level
of machine precision. In fact, to make them visible at all, we had
to multiply the momentum fluxes by a factor $10^{13}$ relative to the
old model.
In other words:  the residual fluxes are controlled by machine accuracy.
Even at this level one cannot detect a correlation between the
fluxes and the position of the sphere.  We can conclude that the
proposed model eliminates the appearance of spurious equilibrium
fluxes.

\section{Conclusions and discussion.}
We have introduced a new model to simulate
the  collective dynamics of non-ideal fluid mixtures, with special
emphasis on its use to study electrokinetic phenomena. The method
relies on a lattice-Boltzmann model, where the interactions are
introduced as effective forces.  In this respect, our model
resembles a Vlasov kinetic model, as opposed to previous kinetic
lattice models. In our approach the fluxes between neighboring
lattice nodes are the fundamental dynamical objects that couple
external fields to both electrical conduction and  hydrodynamic
flow.

As a result of the symmetric formulation of the flux between
neighboring nodes we can impose strict local mass conservation. As
a consequence, the present model is free of spurious boundary
fluxes that plague all other lattice-Boltzmann models of fluid
mixtures. Moreover, a link-based description  has the additional
advantage that boundary conditions are easily implemented.

Secondly, by using a multi-step approach, we can vary ionic
mobilities over many orders of magnitude. This feature of our
model allows us to explore electroviscous effects over a wide
range of Peclet numbers. We have shown that flow causes  spurious
advection-diffusion. However, this effect is well understood and
can be made negligible in most practical cases.

We have checked the performance of the model by studying equilibrium  diffuse layers,
 showing that it is possible to recover  both
 low and high charge density regimes. In the latter, the only
limitation is related to computational resources, because a finer
grid is required to resolve the narrower charge profiles that
develop nearly highly charged walls.
To test the coupling of electrostatics and fluid flow, we have computed the sedimentation
velocity of a charged sphere. These simulations indicate that the
existing theoretical predictions are reproduced in the low-charge,
low-density limit. As the charge of the colloid is increased, the
simulation results start to deviate from the theoretical
predictions that apply in the linearized Poisson-Boltzmann regime.

Even though in the present paper we have focused on electrostatic
interactions and, in particular, we have not discussed molecular
interactions that favor demixing, such interactions could
also be incorporated in the present model.

\section*{Acknowledgments}
This work is part of the research program of the "Stichting voor
Fundamenteel Onderzoek der Materie (FOM)", which is financially
supported by the "Nederlandse organisatie voor Wetenschappelijk
Onderzoek (NWO)". I.P. acknowledges  partial financial support
from DGICYT of the Spanish Government, and
thanks the FOM Institute for its hospitality.

\clearpage

\begin{figure}
\includegraphics[width=0.9\textwidth]{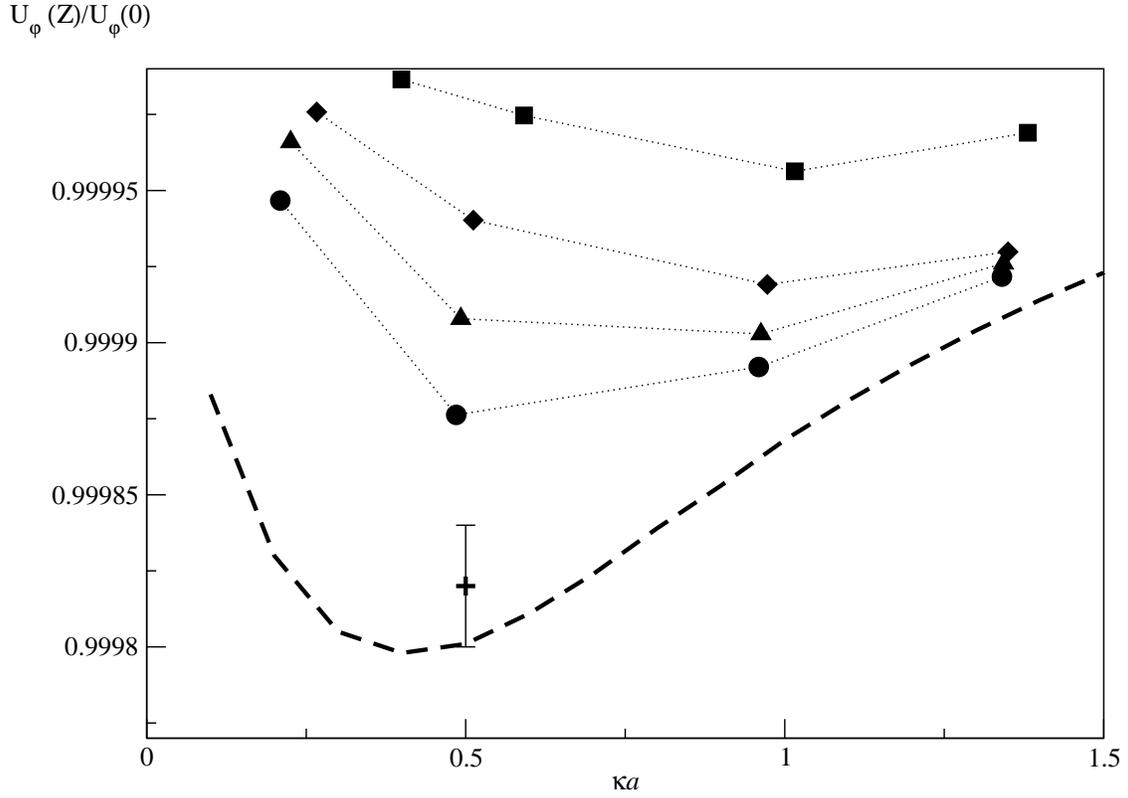}

\caption{Sedimentation velocity of a periodic array of spheres of valency
$Z=10$ and hydrodynamic radius $a= 4.3$. The Bjerrum length
$l_B=0.4$ (in lattice units). The diffusion coefficient of both
positive and negative ions is set to $D=0.19$. We compare
simulation results for finite volume fractions, namely 0.0416
(squares), 0.0123 (diamonds), 0.00521 (triangles), and 0.00267
(circles) against the Booth theory, which is valid at infinite
dilution (dashed curve). For \protect\( \kappa a=0.5\protect \) we
also show the estimated value of the sedimentation velocity at
infinite dilution (see text). The point corresponds to the
extrapolation of the law Eq.~(\ref{U charged dilute limit}).
Within the estimated error, the extrapolated simulation results
agree with the predictions of ref.~\cite{Booth}.
\label{fig5:booth
test} }
\end{figure}

\begin{figure}
\subfigure{
\includegraphics[width=0.6\textwidth]{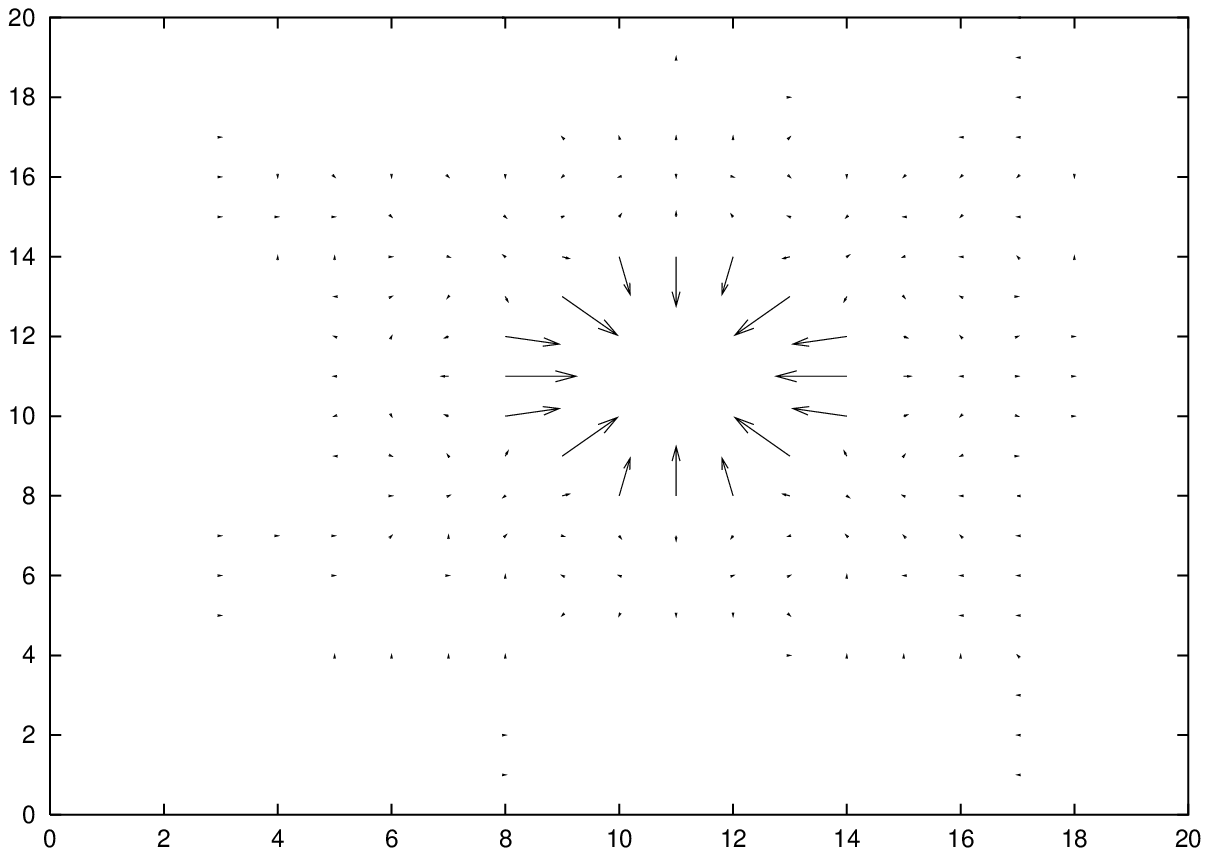}
}
\subfigure{
\includegraphics[width=0.6\textwidth]{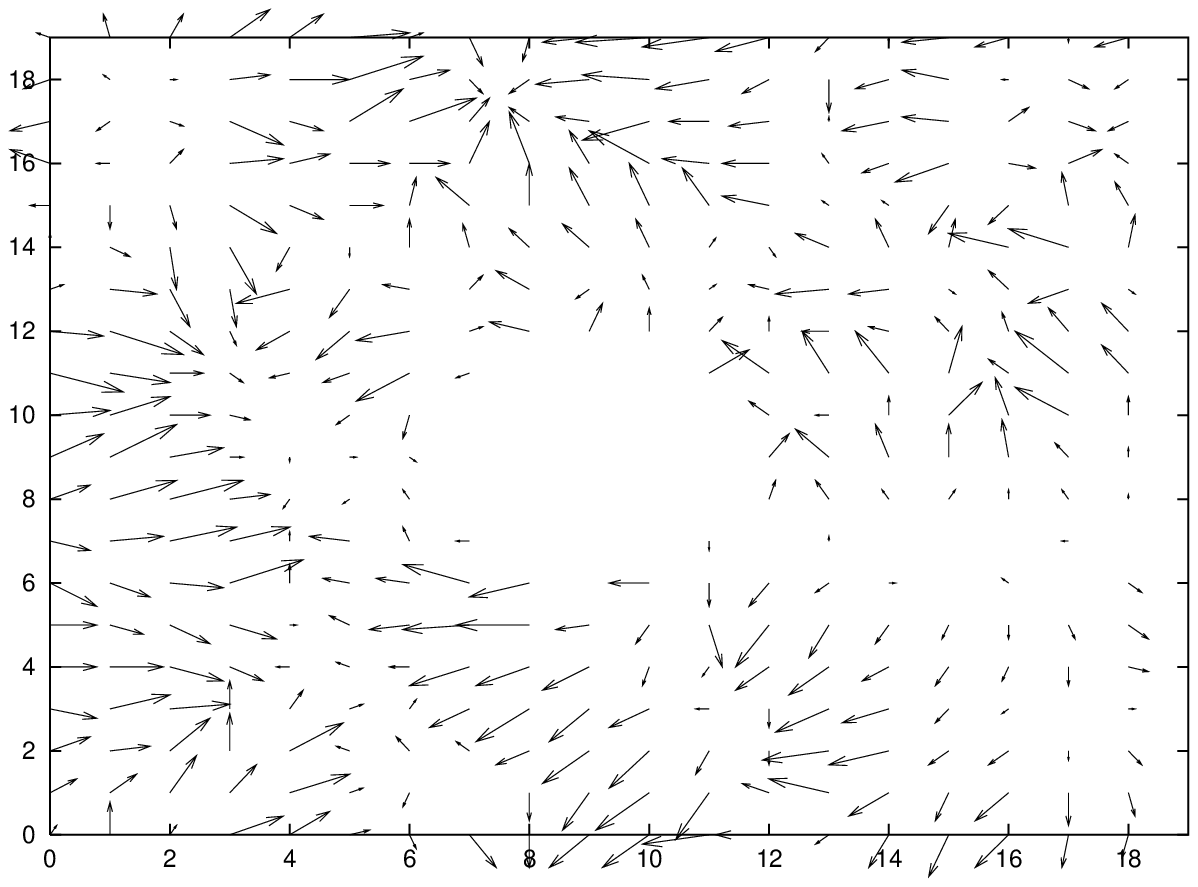}
}

\caption{Illustration of suppression of spurious boundary currents
in present LB model. In the figure we compare the apparent
currents in equilibrium for two models: figure a) gives the
results for the model described in ref.~\cite{Juergen}, figure b)
shows the results for the present model.  In both cases we
consider a colloidal sphere of radius $2.5$ in a system with a
diameter $L=20$ lattice spacings. As there are no external forces
acting on the system and the colloid is not moving, the fluid is
supposed to be at rest. The figure shows the measured projection
of the momentum flux in the equatorial plane of the colloid.
In figure a), spurious currents are apparent close to the particle
surface. The spurious currents in case b) are much smaller than in
case a). In fact, to make them visible at all, they have been
scaled up by a factor $10^{13}$ with respect to case a). This is
an indication that the spurious currents in case b) have been
suppressed down to machine accuracy.
\label{fig6:spurious currents}}
\end{figure}


\begin{thebibliography}{999}
\bibitem{holm} {\sl  Electrostatic effects in Soft Matter and Biophysics}, C. Holm, P. K\`ekicheff and R. Podgornik eds., NATO Sci. Series, (Kluwer Ac. Pub., Dordrecht, 2001)
\bibitem{pincus} W.M. Gelbart, R.F. Bruinsma and P.A. Pincus, Phys. Today {\bf 53}, 38 (2000).
\bibitem{Probstein} R.F. Probstein, {\sl Physicochemical Hydrodynamics} (Butterworth, Boston, 1989).
\bibitem{stone} A.D. Stroock, S.K.W. Deringer, A. Ajdari, I. Mezic, H.A. Stone and G.M. Whitesides, Science {\bf 295}, 647 (2002).
\bibitem{groot} R.D. Groot, J. Chem. Phys, {\bf 118}, 11265 (2003).
\bibitem{PBWarren} P.B. Warren, Int. J. Modern Phys. C {\bf 8}, 889 (1997).
\bibitem{Juergen} J. Horbach and D. Frenkel, Phys. Rev. E {\bf 64} 061507 (2001).
\bibitem{swift} W.R. Osborn, E. Orlandini, M.R. Swift and J.M. Yeomans, Phys. Rev. Lett. {\bf 75}, 4031 (1995);
M.R. Swift, E. Orlandini, W.R. Osborn and J.M. Yeomans, Phys. Rev.
E {\bf 54}, 5041 (1996).
\bibitem{dGM} S.R. de Groot and P. Mazur, {\sl Non-equilibrium thermodynamics} (Dover, New York, 1984).
\bibitem{succi} S. Succi, {\sl The Lattice Boltzmann Equation for Fluid Dynamics and Beyond}, (University Press, Oxford, 2001);
R. Benzi, S. Succi and M. Vergassola, Phys. Rep. , {\bf 222}, 145 (1992).
\bibitem{LaddJSP} A.J.C. Ladd and R. Verberg, J. Stat. Phys. {\bf 104}, 1191 (2001).
\bibitem{guo} Z. Guo, C. Zheng and B. Shi, Phys. Rev. E {\bf 65}, 046308 (2002).
\bibitem{luo} L.-S. Luo and S.S. Girimaji, Phys. Rev. E {\bf 66}, 035301 (2002)
\bibitem{LaddJFMI} A.J.C. Ladd, J. Fluid Mech. {\bf 271}, 285 (1994).
\bibitem{ngai} N.-Q. Nguyen and A.J.C. Ladd, Phys. Rev. E {\bf 66}, 046708 (2002).
\bibitem{Numerical Recipes} W.H. Press, S.A.
Teukolski and W.T. Vetterling, {\sl Numerical Recipes} (Cambridge
University Press, Cambridge, 1996).
\bibitem{Israelachvili} J.N. Israelachvili, {\sl Intermolecular and Surface Forces} (Academic Press, New York, 1991).
\bibitem{Booth} F. Booth, J. Chem. Phys. {\bf 22}, 1956 (1954).
\bibitem{LaddJFMII} A.J.C. Ladd, J. Fluid Mech. {\bf 271}, 311 (1994).
\bibitem{Hashimoto} H. Hashimoto, J Fluid Mech. {\bf 5}, 317 (1959).
\bibitem{LaddJCP} A.J.C. Ladd, J. Chem. Phys. {\bf 93}, 3484 (1990).
\end{thebibliography}
\end{document}